\documentclass[11pt]{article}
\begin{document}
\title{{\bf{\Large Quantum Tunneling and Back Reaction }}}
\author{
 {\bf {\normalsize Rabin Banerjee}$
$\thanks{E-mail: rabin@bose.res.in}},\, 
 {\bf {\normalsize Bibhas Ranjan Majhi}$
$\thanks{E-mail: bibhas@bose.res.in}}\\
 {\normalsize S.~N.~Bose National Centre for Basic Sciences,}
\\{\normalsize JD Block, Sector III, Salt Lake, Kolkata-700098, India}
\\[0.3cm]
}

\maketitle

{\bf Abstract:}\\
We give a correction to the tunneling probability by taking into account the back reaction effect to the metric of the black hole spacetime. We then  show how this gives rise to the modifications in the semiclassical black hole entropy and Hawking temperature. Finally, we reproduce the familiar logarithmic correction to the Bekenstein-Hawking area law.\\

    In 1975, Hawking discovered \cite{Hawking} the remarkable fact that black holes, previously thought to be completely black regions of spacetime from which nothing can escape, actually radiate a thermal spectrum of paticles and that the temperature of this radiation depends on the surface gravity of the black hole by the relation $T_H=\frac{{\cal{K}}_{0}}{2\pi}$. This discovery was consistent with an earlier discovery \cite{Bardeen,Bekenstein1,Bekenstein,Bekenstein2} of a connection between black holes and thermodynamics which revealed that the entropy of a black hole is proportional to the surface area of its horizon, 
$S_{BH}=\frac{A}{4}$. From this, using the second law of thermodynamics $dM=T_HdS_{BH}$, the temperature of the black hole can be calculated. This is actually due to an analogy between the second law of thermodynamics and the black hole equation $dM=\frac{{\cal{K}}_{0}}{8\pi}dA$. These relations are all based on classical or semiclassical considerations.

   It is possible to include quantum effects in this discussion of Hawking radiation. Using the conformal anomaly method the modifications to the spacetime metric by the one loop back reaction was computed \cite{York,Lousto}. Later it was shown \cite{Fursaev,Mann} that the Bekenstein-Hawking area law was modified, in the leading order, by logarithmic corrections. Similar conclusions were also obtained by using quantum gravity techniques \cite{Parthasarathi1,Mitra,Page}. Likewise, corrections to the semiclassical Hawking temperature were derived \cite{Lousto,Fursaev}. 

     A particularly useful and intuitive way to understand the Hawking effect is through the tunneling formalism as developed in \cite{Wilczek}. The semiclassical Hawking temperature is very simply and quickly obtained \cite{Wilczek,Paddy} in this scheme by exploiting the form of the semiclassical tunneling rate. A natural question that arises in this context is the feasibility of this approach to include quantum corrections. Although there have been sporadic attempts \cite{Medved1,Medved2} a systematic, thorough and complete analysis is still lacking.

      In this paper we compute the corrections to the semiclassical tunneling rate by including the effects of self gravitation and back reaction. The usual expression found in \cite{Wilczek}, given in the Maxwell-Boltzmann form $e^{-\frac{E}{T_{BH}}}$, is modified by a prefactor. This prefactor leads to a modified Bekenstein-Hawking entropy. The semiclassical Bekenstein-Hawking area law connecting the entropy to the horizon area is altered. As obtained in other approaches \cite{Mann,Parthasarathi1,Page}, the leading correction is found to be logarithmic while the nonleading one is a series in inverse powers of the horizon area (or Bekenstein-Hawking entropy). We also compute the appropriate modification to the Hawking temperature. Explicit results are given for the Schwarzschild black hole.\\

      Let us consider a general class of static, spherically symmetric spacetime of the form
\begin{eqnarray}
ds^2 = -f(r)dt^2+\frac{dr^2}{g(r)}+r^2 d\Omega^2
\label{1.01}
\end{eqnarray}
where the horizon $r=r_h$ is given by $f(r_h)=g(r_h)=0$. The metric has a coordinate singularity at the horizon which is removed by transforming to Painleve coordinates \cite{Painleve}. A suitable choice is $dt\to dt-\sqrt{\frac{1-g(r)}{f(r)g(r)}}dr$ under which the metric takes the form, 
\begin{eqnarray}
ds^2 = -f(r)dt^2+2f(r)\sqrt{\frac{1-g(r)}{f(r)g(r)}} dt dr+dr^2+r^2d\Omega^2
\label{1.02}
\end{eqnarray}
The radial null geodesics ($ds^2=d\Omega^2=0$) are given by
\begin{eqnarray}
\dot{r}\equiv\frac{dr}{dt}=\sqrt{\frac{f(r)}{g(r)}}\Big(\pm 1-\sqrt{1-g(r)}\Big)
\label{1.03}
\end{eqnarray}
where the positive (negative) sign gives outgoing (incoming) radial geodesics. Now expanding the quantitis $f(r)$ and $g(r)$ about the horizon $r_h$ we get
\begin{eqnarray}
&&f(r)=f'(r_h)(r-r_h)+O((r-r_h)^2)
\nonumber
\\
&&g(r)=g'(r_h)(r-r_h)+O((r-r_h)^2)
\label{1.04}
\end{eqnarray}
The surface gravity of the black hole on the horizon is defined by
\begin{eqnarray}
{\cal{K}}(M)=\Gamma{^0}{_{00}}|_{r=r_h}=\frac{1}{2}\Big[\sqrt{\frac{1-g(r)}{f(r)g(r)}}g(r)\frac{df(r)}{dr}\Big]|_{r=r_h}
\label{1.05}
\end{eqnarray}
Therefore using (\ref{1.04}) ${\cal{K}}$ and $\dot{r}$ can be approximately  expressed as
\begin{eqnarray}
{\cal{K}}(M)\simeq\frac{1}{2}\sqrt{f'(r_h)g'(r_h)}
\label{1.06}
\end{eqnarray}
and
\begin{eqnarray}
\dot{r}\simeq&&\frac{1}{2}\sqrt{f'(r_h)g'(r_h)}(r-r_h)+O((r-r_h)^2)
\nonumber
\\
&&={\cal{K}}(M)(r-r_h)+O((r-r_h)^2)
\label{1.07}
\end{eqnarray}
where in the last step (\ref{1.06}) has been used.

 The imaginary part of the action for an $s$-wave outgoing positive energy particle which crosses the horizon outwards from $r_{in}$ to $r_{out}$ can be expressed as,
\begin{eqnarray}
Im {\cal{S}} =Im \int_{r_{in}}^{r_{out}} p_r dr = Im  \int_{r_{in}}^{r_{out}} \int_{0}^{p_r} dp_r'dr = Im \int_{r_{in}}^{r_{out}} \int_{0}^{H} \frac{dH'}{\dot{r}} dr
\label{1.1}
\end{eqnarray}
where in the last step we multiply and divide the integrand by the two sides of Hamilton's equation $\dot{r}=\frac{dH}{dp_r}|_r$. Now taking into account the self-gravitation effects \cite{Kraus}, the above integration can be expressed as,  
\begin{eqnarray}
Im {\cal{S}} = Im \int_{r_{in}}^{r_{out}} \int_{M}^{M-\omega} \frac{dH'}{\dot{r}} dr  =-Im \int_{r_{in}}^{r_{out}} \int_{0}^{\omega} \frac{d\omega'}{\dot{r}} dr
\label{1.2}
\end{eqnarray}
where we used the fact that the Hamiltonian $H=M-\omega$, with $M$ being the original mass of the black hole. Here $\dot{r}$ can be approximated by (\ref{1.07}) as follows,
\begin{equation}
\dot{r}\simeq (r-r_h){\cal{K}}(M-\omega)+O((r-r_h)^2)
\label{1.3}
\end{equation}
where $r_h$ is the modified Schwarzschild radius and ${\cal{K}}(M-\omega)$ is the modified horizon surface gravity. This modification occures due to two effects; self-gravitation which requires the replacement $M\rightarrow M-\omega$ so that ${\cal{K}}(M)\rightarrow{\cal{K}}(M-\omega)$ and back reaction to be discussed below. Taking only the first order term of $\dot{r}$, the integration in (\ref{1.2}) can be written as,
\begin{eqnarray}
Im {\cal{S}} =  -Im \int_{r_{in}}^{r_{out}} \int_{0}^{\omega} \frac{d\omega'}{(r-r_h){\cal{K}}(M-\omega')} dr
\label{1.4}
\end{eqnarray}
But now the integration over $r$ can be done by deforming the contour. Ensuring that the positive energy solutions decay in time (i.e. into the lower half of $\omega'$ plane and $r_{in}>r_{out}$) we have after $r$ integration {\footnote{A similar treatment can be done by taking the contour in the upper half plane but then one has to replace $M\rightarrow M+\omega$ \cite{Kraus}.}}, 
\begin{eqnarray}
Im {\cal{S}}= \pi\int_{0}^{\omega} \frac{d\omega'}{{\cal{K}}(M-\omega')} 
\label{1.5}
\end{eqnarray}

    A derivation of (\ref{1.5}), to the leading order in $\omega$, following similar techniques has been presented in \cite{Ang,Ryan}. In fact if we expand $\frac{1}{{\cal{K}}(M-\omega')}$ retaining terms linear in $\omega'$, we immediately find,
\begin{eqnarray}
Im {\cal{S}}&=& \pi\int_{0}^{\omega} {d\omega'}\Big[\frac{1}{{\cal{K}}(M)}+{\cal{O}}(\omega')\Big] 
\nonumber
\\
&=&\frac{\pi\omega}{{\cal{K}}(M)}+{\cal{O}}(\omega^2)
\nonumber
\\
&=&\frac{2\pi\omega}{\sqrt{f'(r_h)g'(r_h)}}+{\cal{O}}(\omega^2)
\label{1.500}
\end{eqnarray}   
which is the expression given in \cite{Ang,Ryan}. In getting the final form the value of ${\cal{K}}(M)$ from (\ref{1.06}) has been used.

Now the modified surface gravity due to one loop back reaction effects is given by \cite{York,Lousto},
\begin{eqnarray}
{\cal{K}}(M) = {\cal{K}}_0(M)\Big(1+\frac{\alpha}{M^2}\Big)
\label{1.6}
\end{eqnarray}
where ${\cal{K}}_0$ is the classical surface gravity at the horizon of the black hole. Such a form is dictated by simple scaling arguments. As is well known, a loop expansion is equivalent to an expansion in powers of the Planck constant $\hbar$. Since, in natural units, $\sqrt{\hbar}=M_p$, the one loop correction has a form given by $\frac{\alpha}{M^2}$. The constant $\alpha$ is related to the trace anomaly coefficient taking into account the degrees of freedom of the fields \cite{Lousto,Fursaev}. Its explicit form is given by \cite{Fursaev}, 
\begin{eqnarray}
\alpha=\frac{1}{360\pi}\Big(-N_0 - \frac{7}{4}N_{\frac{1}{2}} +13N_1 + \frac{233}{4}N_{\frac{3}{2}} - 212 N_2) 
\label{1.600}
\end{eqnarray} 
where $N_s$ denotes the number of fields with spin `$s$'.

    For the classical Schwarzschild spacetime 
\begin{eqnarray}
f(r)=g(r)=  1-\frac{2M}{r};\,\,\,r_H=2M
\label{new3}
\end{eqnarray}
and so by equation (\ref{1.06})
the value of ${\cal{K}}_0(M)$ is
\begin{eqnarray}
{\cal{K}}_0(M) = \frac{f'(r_H=2M)}{2}=\frac{1}{4M}
\label{1.61}
\end{eqnarray}
Substituting these in (\ref{1.5}) and then integrating over $\omega'$ we have 
\begin{eqnarray}
Im {\cal{S}} = 4\pi\omega(M-\frac{\omega}{2})+2\pi\alpha \ln{\Big[\frac{(M-\omega)^2+\alpha}{M^2+\alpha}\Big]}
\label{1.7}
\end{eqnarray}
Now according to the WKB-approximation method the tunneling probability is given by,
\begin{eqnarray}
\Gamma\sim e^{-2 Im {\cal{S}}} 
\label{1.8}
\end{eqnarray}
So the modified tunneling probability due to back reaction effects is, 
\begin{eqnarray}
\Gamma\sim \Big[1-\frac{2\omega(M-\frac{\omega}{2})}{M^2+\alpha}\Big]^{-4\pi\alpha} e^{-8\pi\omega(M-\frac{\omega}{2})} 
\label{1.9}
\end{eqnarray}
The exponential factor of the tunneling probability was previously obtained by Parikh and Wilczek \cite{Wilczek}. The factor before the exponential is actually due the effect of back reaction. It will eventually give the correction to the Bekenstein-Hawking entropy and the Hawking temperature as will be shown below.

    It is known \cite{Wilczek,Pilling,Sarkar} that change in the Bekenstein-Hawking entropy due to the tunneling through the horizon is related to $Im {\cal{S}}$ by the following relation,  
\begin{eqnarray}
\Delta S_{bh} = -2Im {\cal{S}} 
\label{1.10}
\end{eqnarray}
Therefore the corrected change in Bekenstein-Hawking entropy is 
\begin{eqnarray}
\Delta S_{bh}=-8\pi\omega(M-\frac{\omega}{2})-4\pi\alpha\ln\Big[(M-\omega)^2+\alpha\Big]+4\pi\alpha\ln(M^2+\alpha)
\label{1.11}
\end{eqnarray}
Next using the stability criterion $\frac{d(\Delta S_{bh})}{d\omega}=0 $ for the black hole, one obtains the following condition
\begin{eqnarray}
(\omega-M)^3=0 
\label{1.12}
\end{eqnarray}
which gives the only solution as $\omega=M$. Substituting this value of $\omega$ in (\ref{1.11}) we will have the change in entropy of the black hole from its initial state to final state.
\begin{eqnarray}
S_{final}-S_{initial}= -4\pi M^2+4\pi\alpha\ln{(\frac{M^2}{\alpha}+1)}
\label{1.13}
\end{eqnarray}
So the Bekenstein-Hawking entropy of the black hole with mass $M$ is 
\begin{eqnarray}
S_{bh}&=& 4\pi M^2-4\pi\alpha\ln{(\frac{M^2}{\alpha}+1)}
\label{new1}
\end{eqnarray}
Ignoring back reaction (i.e. $\alpha=0$) we just reproduce the usual semiclassical area law \cite{Bardeen,Bekenstein1,Bekenstein,Bekenstein2} for the Bekenstein-Hawking entropy,
\begin{eqnarray}
S_{BH} =4\pi M^2=\frac{A}{4}
\label{1.15}
\end{eqnarray}
where $A$ is the area of the black hole horizon given by,
\begin{eqnarray}
A=4\pi r^2_H=16\pi M^2
\label{new4}
\end{eqnarray}
Substituting (\ref{1.15}) in (\ref{new1}) and expanding the logarithm, we obtain the final form,
\begin{eqnarray}
S_{bh}&=&\frac{A}{4}-8\pi\alpha\ln{M}-64\pi^2\alpha^2\Big[\frac{1}{A}-\frac{16\pi\alpha}{2A^2}+\frac{(16\pi\alpha)^2}{3A^3}-.....\Big]
\nonumber
\\
&+&\textrm{const.(independent~ of~ M)}
\nonumber
\\
&=&S_{BH}-4\pi\alpha\ln{S_{BH}}-\frac{16\pi^2\alpha^2}{S_{BH}}\Big[1-\frac{(4\pi\alpha)}{2S_{BH}}+\frac{(4\pi\alpha)^2}{3(S_{BH})^2}-.....\Big]
\nonumber
\\
&+&\textrm{const.(independent~ of~ M)}
\label{1.14}
\end{eqnarray}
The well known logarithmic correction \cite{Fursaev,Mann,Parthasarathi1,Mitra,Page} appears in the leading term. Quantum gravity calculations lead to a prefactor $-\frac{1}{2}$ for the $\ln S_{BH}$ term which would correspond to choosing $\alpha=\frac{1}{8\pi}$. Also, the nonleading corrections are found to be expressed as a series in inverse powers of $A$(or $S_{BH}$), exactly as happens in quantum gravity inspired analysis \cite{Parthasarathi1,Page}.

     Now using the second law of thermodynamics 
\begin{eqnarray}
T_h dS_{bh}=dM
\label{new2}
\end{eqnarray}
we can find the corrected form of the Hawking temperature $T_h$ due to back reaction. This is obtained from (\ref{new1}) 
\begin{eqnarray}
\frac{1}{T_h}=\frac{dS_{bh}}{dM}=8\pi M\Big(\frac{M^2}{M^2+\alpha}\Big)
\label{1.16}
\end{eqnarray}
Therefore the corrected Hawking temperature is given by
\begin{eqnarray}
T_h = T_H\Big(1+\frac{\alpha}{M^2}\Big) 
\label{1.17}
\end{eqnarray}
where $T_H = \frac{1}{8\pi M}$ is the semiclassical Hawking temperature and the other term is the correction due to the back reaction. A similar expression was obtained previously in \cite{Fursaev} by the conformal anomaly method.

   It is also possible to obtain the corrected Hawking temperature (\ref{1.17}) in the standard tunneling method to leading order \cite{Wilczek} where this temperature is read off from the coefficient of `$\omega$' in the exponential of the probability amplitude (\ref{1.9}). Recasting this amplitude as,
\begin{eqnarray}
\Gamma\sim e^{-8\pi\omega(M-\frac{\omega}{2})-4\pi\alpha \ln(1-\frac{2\omega(M-\frac{\omega}{2})}{M^2+\alpha})} 
\end{eqnarray} 
and retaining terms upto leading order in $\omega$, we obtain,
\begin{eqnarray}
\Gamma &\sim& e^{-8\pi M \omega+4\pi\alpha (\frac{2M\omega}{M^2+\alpha})}
\nonumber
\\
&=&e^{-(\frac{8\pi M^3}{M^2+\alpha})\omega}=e^{-\frac{\omega}{T_h}}. 
\end{eqnarray}
The inverse Hawking temperature, indentified with the coefficient of `$\omega$',
reproduces (\ref{1.17}).

    It is also observed that the usual (semiclassical) identification between the surface gravity and Hawking temperature $T_H=\frac{{\cal{K}}_0(M)}{2\pi}$ persists even after including the back reaction. From (\ref{1.6}) and (\ref{1.17}) we easily infer that $T_h=\frac{{\cal{K}}(M)}{2\pi}$.\\

   To conclude, we have considered self-gravitation and (one loop) back reaction effects in tunneling formalism for Hawking radiation. The modified tunneling rate was computed. From this modification, corrections to the semiclassical expressions for  entropy and Hawking temperature were obtained. Also, the logarithmic corrections to the semiclassical Bekenstein-Hawking area law was reproduced. Although our analysis was presented for the Schwarzschild black hole, it is general enough to include other examples as well.

\end{document}